\documentstyle[12pt]{article}
\def\hybrid{\topmargin -20pt    \oddsidemargin 0pt
        \headheight 0pt \headsep 0pt
        \textwidth 6.25in       % A4 paper
        \textheight 9.5in       % A4 paper
        \marginparwidth .875in
        \parskip 5pt plus 1pt   \jot = 1.5ex}

\hybrid

\def\baselinestretch{1.2}

\catcode`\@=11

\def\marginnote#1{}
\newcount\hour
\newcount\minute
\newtoks\amorpm
\hour=\time\divide\hour by60
\minute=\time{\multiply\hour by60 \global\advance\minute by-\hour}
\edef\standardtime{{\ifnum\hour<12 \global\amorpm={am}%
        \else\global\amorpm={pm}\advance\hour by-12 \fi
        \ifnum\hour=0 \hour=12 \fi
        \number\hour:\ifnum\minute<10 0\fi\number\minute\the\amorpm}}
\edef\militarytime{\number\hour:\ifnum\minute<10 0\fi\number\minute}
\def\draftlabel#1{{\@bsphack\if@filesw {\let\thepage\relax
   \xdef\@gtempa{\write\@auxout{\string
      \newlabel{#1}{{\@currentlabel}{\thepage}}}}}\@gtempa
   \if@nobreak \ifvmode\nobreak\fi\fi\fi\@esphack}
        \gdef\@eqnlabel{#1}}
\def\@eqnlabel{}
\def\@vacuum{}
\def\draftmarginnote#1{\marginpar{\raggedright\scriptsize\tt#1}}

\def\draft{\oddsidemargin -.5truein
        \def\@oddfoot{\sl preliminary draft \hfil
        \rm\thepage\hfil\sl\today\quad\militarytime}
        \let\@evenfoot\@oddfoot \overfullrule 3pt
        \let\label=\draftlabel
        \let\marginnote=\draftmarginnote
   \def\@eqnnum{(\theequation)\rlap{\kern\marginparsep\tt\@eqnlabel}%
\global\let\@eqnlabel\@vacuum}  }

\def\preprint{\twocolumn\sloppy\flushbottom\parindent 2em
        \leftmargini 2em\leftmarginv .5em\leftmarginvi .5em
        \oddsidemargin -.5in    \evensidemargin -.5in
        \columnsep .4in \footheight 0pt
        \textwidth 10.in        \topmargin  -.4in
        \headheight 12pt \topskip .4in
        \textheight 6.9in \footskip 0pt
        \def\@oddhead{\thepage\hfil\addtocounter{page}{1}\thepage}
        \let\@evenhead\@oddhead \def\@oddfoot{} \def\@evenfoot{} }

\def\numberbysection{\@addtoreset{equation}{section}
        \def\theequation{\thesection.\arabic{equation}}}

\def\underline#1{\relax\ifmmode\@@underline#1\else
        $\@@underline{\hbox{#1}}$\relax\fi}

\def\titlepage{\@restonecolfalse\if@twocolumn\@restonecoltrue\onecolumn
     \else \newpage \fi \thispagestyle{empty}\c@page\z@
        \def\thefootnote{\fnsymbol{footnote}} }

\def\endtitlepage{\if@restonecol\twocolumn \else \newpage \fi
        \def\thefootnote{\arabic{footnote}}
        \setcounter{footnote}{0}}  %\c@footnote\z@ }

\catcode`@=12
\relax

\def\figcap{\section*{Figure Captions\markboth
        {FIGURECAPTIONS}{FIGURECAPTIONS}}\list
        {Figure \arabic{enumi}:\hfill}{\settowidth\labelwidth{Figure
999:}
        \leftmargin\labelwidth
        \advance\leftmargin\labelsep\usecounter{enumi}}}
 \relax
\def\tablecap{\section*{Table Captions\markboth
        {TABLECAPTIONS}{TABLECAPTIONS}}\list
        {Table \arabic{enumi}:\hfill}{\settowidth\labelwidth{Table
999:}
        \leftmargin\labelwidth
        \advance\leftmargin\labelsep\usecounter{enumi}}}
 \relax
\def\reflist{\section*{References\markboth
        {REFLIST}{REFLIST}}\list
        {[\arabic{enumi}]\hfill}{\settowidth\labelwidth{[999]}
        \leftmargin\labelwidth
        \advance\leftmargin\labelsep\usecounter{enumi}}}
 \relax
\makeatletter
\newcounter{pubctr}
\def\publist{\@ifnextchar[{\@publist}{\@@publist}}
\def\@publist[#1]{\list
        {[\arabic{pubctr}]\hfill}{\settowidth\labelwidth{[999]}
        \leftmargin\labelwidth
        \advance\leftmargin\labelsep
        \@nmbrlisttrue\def\@listctr{pubctr}
        \setcounter{pubctr}{#1}\addtocounter{pubctr}{-1}}}
\def\@@publist{\list
        {[\arabic{pubctr}]\hfill}{\settowidth\labelwidth{[999]}
        \leftmargin\labelwidth
        \advance\leftmargin\labelsep
        \@nmbrlisttrue\def\@listctr{pubctr}}}
 \relax
\makeatother
\newskip\humongous \humongous=0pt plus 1000pt minus 1000pt

\newif\ifdtup

\relax

\def\be{\begin{equation}}
\def\ee{\end{equation}}
\def\ba{\begin{eqnarray}}
\def\ea{\end{eqnarray}}

\def\del{\partial}

\def\a{\alpha}

\def\b{\beta}

\def\e{\epsilon}

\def\th{\theta}

\def\m{\mu}
\def\n{\nu}
\def\om{\omega}

\def\no{\noindent}

\def\qq{\qquad}

\def\IR{\relax{\rm I\kern-.18em R}}

\def \ov {\over}

\def\IR{\relax{\rm I\kern-.18em R}}
\def\inv{^{\raise.15ex\hbox{${\scriptscriptstyle -}$}\kern-.05em 1}}

\def\tL{{\tilde L}}

\begin{document}
%\draft

%\renewcommand{\theequation}{\arabic{equation}}
\renewcommand{\theequation}{\thesection.\arabic{equation}}

\newcommand{\beq}{\begin{equation}}
\newcommand{\eeq}[1]{\label{#1}\end{equation}}
\newcommand{\ber}{\begin{eqnarray}}
\newcommand{\eer}[1]{\label{#1}\end{eqnarray}}
\newcommand{\eqn}[1]{(\ref{#1})}
\begin{titlepage}
\begin{center}

\hfill NEIP-00-021\\
\vskip -.1 cm
\hfill hep-th/0012125\\
\vskip -.1 cm
\hfill December 2000\\

\vskip .4in

{\large \bf Gravitational domain walls and $p$-brane distributions
%{\large \bf Algebraic-curve classification of p-brane distributions
%in M- and string theory 
\footnote{Based on talks presented by I.B. at the 9th Hellenic conference
on Relativity (NEB IX) held in Ioannina, Greece, 28--31 August 2000, 
and at the 1st 
workshop of the
RTN network {\em ``The quantum structure of spacetime and the 
geometric nature of fundamental interactions"} (34th International 
Symposium Ahrenshoop on the Theory of Elementary Particles) held in 
Berlin, Germany, 4--10 October 2000; 
to appear in the proceedings as special issue
of {\em Fortschritte der Physik} (Wiley-VCH)}}

\vskip 0.4in

{\bf Ioannis Bakas${}^1$},\phantom{x} 
and\phantom{x} {\bf Konstadinos Sfetsos${}^2$}
\vskip 0.1in
{\em ${}^1\!$Department of Physics, University of Patras \\
GR-26500 Patras, Greece\\
\footnotesize{\tt bakas@ajax.physics.upatras.gr}}\\
\vskip .2in
{\em ${}^2\!$Institut de Physique, Universit\'e de Neuch\^atel\\
rue Breguet 1, CH-2000 Neuch\^atel, Switzerland\\
\footnotesize
{\tt konstadinos.sfetsos@unine.ch}}\\

\end{center}

\vskip .6in

\centerline{\bf Abstract}

\no
We review the main algebraic aspects that characterize and determine the
domain wall solutions of maximal gauged supergravity in various spacetime
dimensions by considering consistent truncations that retain a 
number of components in the diagonal of the coset space scalar manifold
of the theory. 
Starting from the algebraic classification of domain walls in $D=4$
gauged supergravity, we also derive the corresponding solutions in $D=5$
and $D=7$ dimensions.
>From a higher dimensional point of view, these solutions
describe the gravitatonal field, in the field theory limit,
of a large number of M2-, D3- and M5-branes that are
distributed on hypersurfaces in the transverse space to the branes. 
As a new result we employ a smearing procedure as well as various 
dualities to list the irreducible curves and the symmetry groups of 
$p$-brane distributions for all values of $p$ that are of interest 
in current applications of string theory. 
Some emphasis is placed on the presentation of new
results in the case of NS5-branes.   
\no

\vfill
\end{titlepage}
\eject

\def\baselinestretch{1.2}
\baselineskip 16 pt
\noindent

%%%%%%%%%%%%%%%%%%%%%A generalisation of target space%%%%%%%%%
\def\tT{{\tilde T}}
\def\tg{{\tilde g}}
\def\tL{{\tilde L}}

%%%%%%%%%%%%%%%%%%%%%%%%%%%%%%%%%%%%%%%%%%%%%%%%%%%%%%%%%%%%%%

\section{Gravitational domain walls}

It is well known that gravitational theories can admit domain wall 
solutions in the presence of scalar fields. In particular, $D$-dimensional 
gravity coupled to a scalar field $\phi$ with potential $V(\phi)$ that is 
derived from a superpotential $W(\phi)$, as in 
\be
S[g, \phi]  =  \int d^D x \left({1 \over 4} R - {1 \over 2} 
(\partial \phi)^2 - V(\phi)
\right), \label{act1}
\ee
with
\be
V(\phi)  =  {g^2 \over 8} \left(\left({\partial W \over \partial \phi} 
\right)^2 - 2{D-1 \over D-2} W^2 \right)    \ ,
\label{supp}
\ee
serves as a prototype for constructing 
stable solutions that depend on a single spatial coordinate, say $r$,
using the special ansatz for the spacetime metric
\be
ds^2 = dr^2 + e^{2A(r)}{\eta}_{\mu \nu} dx^{\mu} dx^{\nu} ~; ~~~~ \mu, \nu = 
0, 1, \cdots , D-2 ~, 
\label{mean1}
\ee
which preserves Poincar\'e invariance in $D-1$ dimensions. The domain walls 
satisfy the first order equations
\be
{d \phi \over dr} = \mp {g \over 2} {\partial W \over \partial \phi} ~ , ~~~~~ 
{dA \over dr} = \pm {g \over D-2} W ~,  
\label{doan1}
\ee
which are equivalent conditions for having $1/2$-BPS 
supersymmetric configurations in the bosonic sector of gauged supergravity,
in appropriate generalizations of the model. 
  
Extending the standard Bogomol'nyi argument to the gravitational case, one
is lead to consider the following effective functional
\be
E[A, \phi] = \int_{-\infty}^{+\infty} dr e^{(D-1)A} \left(
{1 \over 2}({\partial}_r \phi)^2 
+ V(\phi) - {1 \over 4}(D-1)(D-2) ({\partial}_r A)^2 \right) \ ,
\label{act2}
\ee
whose extrema provide an equivalent description of all solutions of the
theory \eqn{act1} in the sector with metric form \eqn{mean1}. The special set of
first order equations \eqn{doan1} that characterize the domain walls follow
easily by completing the squares in the integrant of the effective
functional as
\ba
E & = & {1 \over 2} \int_{-\infty}^{+\infty} dr e^{(D-1)A} 
\left(\left({\partial}_r \phi 
\pm {g \over 2}
{\partial W \over \partial \phi} \right)^2  
-  {1 \over 2}(D-1)(D-2) 
\left({\partial}_r A \mp {g \over D-2} W \right)^2 \right) \nonumber\\ 
& \mp & {g \over 2} W e^{(D-1)A} {\mid}_{-\infty}^{+\infty} ~,   
\ea
where $\pm \infty$ represent the end points in the range of the 
variable $r$.
When each one of the two perfect square terms vanishes separately, 
the effective 
functional receives contribution only from the boundary term, and the 
resulting first order equations arise as its saddle points (accounting for
the relative $-$ sign between the two terms); as such they also provide 
solutions of the full set of second order equations.    

We note for completeness that the gravitational domain walls provide a natural 
generalization of the usual  
kink solutions arising in the 2-dim scalar field theory
\be
S[\phi] = \int d^2 x \left( -{1 \over 2} (\partial \phi)^2 - V(\phi) \right) ~; 
~~~~ V(\phi) = {g^2 \over 8} \left({\partial W \over \partial \phi} \right)^2  \ ,
\label{hg1}
\ee
which are 
obtained by minimizing the energy functional of its static configurations a la
Bogomol'nyi
\be
E[\phi] = \int_{-\infty}^{+\infty} dx \left({1 \over 2}({\partial}_x \phi)^2 + 
V(\phi) \right) = {1 \over 2} \int_{-\infty}^{+\infty} dx \left({\partial}_x 
\phi \pm {g \over 2} {\partial W \over \partial \phi}\right)^2 \mp {g \over 2}
W {\mid}_{-\infty}^{+\infty} ~. 
\label{hg2}
\ee
In fact, it can be seen that \eqn{hg1} and \eqn{hg2} follow from the 
corresponding equations in the gravitational case, but in the limit where 
gravity decouples. Indeed, in \eqn{act1} above
we have set the gravitational coupling
constant $\kappa_D=\sqrt{2}$, but if we reinstate it and 
perform a rescaling of the scalar fields, gravity will decouple in the
limit $\kappa_D\to 0$, as stated.

Some basic facts, as well as various applications of gravitational domain 
walls, can be found in the review \cite{cvetic} (and references therein).
The results we present here are based on earlier work 
in \cite{basfe}; also some closely related results can be found in 
\cite{others} (and references therein).
 
\section{Solutions of gauged supergravity}
\setcounter{equation}{0}

In order to construct domain wall solutions of gauged supergravity in various
dimensions, it is first convenient to extend the formalism to a multi scalar
field context with diagonal target space metric, namely
\be
S[g, \phi_I]  = \int d^D x \left({1 \over 4} R - {1 \over 2} 
\sum_{I=1}^{N-1} (\partial \phi_I)^2 - 
{g^2 \over 8} \left(\sum_{I=1}^{N-1} \left({\partial W \over \partial \phi_I} 
\right)^2 - 2{D-1 \over D-2} W^2 \right) \right)    \ .
\ee
Then, the corresponding system of first order equations turns out to be
\be
{d \phi_I \over dr} = {g \over 2} {\partial W \over \partial \phi_I} ~ , ~~~~~ 
{dA \over dr} = - {g \over D-2} W ~,  
\ee
where one of the two signs has been chosen 
for simplicity and $g$ will be set equal to 1 from now on, unless 
otherwise stated.

We consider the sector of maximally gauged supergravity in various dimensions
in which the scalar field 
manifold assumes the coset space form $SL(N, \IR)/SO(N)$. A further consistent truncation 
is to confine ourselves to the
diagonal components, setting all other scalar and gauge fields equal to zero.
As we will see shortly, 
this particular sector turns out to be exactly solvable by employing techniques
of algebraic geometry. It is convenient to introduce $N$ scalar fields 
$\beta_i$ subject to the constraint
\be
\beta_1 + \beta_2 + \cdots + \beta_N = 0 ~, 
\label{constr}
\ee
which are related to the original fields $\phi_I$ as in
\be
\b_i = \sum_{I=1}^{N-1} {\lambda}_{iI} \phi_I \ ,
\ee
with ${\lambda}_{iI}$ being the fundamental weights of $SL(N)$ algebra, which
satisfy the defining relations
\be
\sum_{i=1}^{N} {\lambda}_{iI} = 0 ~, ~~~~ \sum_{i=1}^{N} {\lambda}_{iI}
{\lambda}_{iJ} = 2 {\delta}_{IJ} ~, ~~~~ \sum_{I=1}^{N-1} {\lambda}_{iI}
{\lambda}_{jI} = 2{\delta}_{ij} - {2 \over N} ~. 
\ee
Using this data, the superpotential of gauged supergravity assumes the
simple form
\be
W = -{1 \over 4} \sum_{i=1}^{N} e^{2 \beta_i} ~.
\ee
 
We will construct domain wall equations of the theory using the conformally
flat form of the metric
\be
ds^2 = e^{2A(z)} \left(dz^2 + {\eta}_{\mu \nu} dx^{\mu} dx^{\nu} \right) , 
\label{mean2}
\ee
or else the coordinate $z$, instead of $r$, related as $({\rm exp}A) dz = -dr$. 
Writing the first order equations for the 
domain walls, we obtain the following system in 
terms of the set of fields $\beta_i$:
\be
{\beta}_i^{\prime} = {1 \over 2} e^{2 \beta_i + A} + 2{D-2 \over N} A^{\prime}
~, ~~~~ A^{\prime} = {1 \over D-2} e^{A} W \ ,
\label{baeq}
\ee
with $i = 1, 2, \cdots , N$, and prime denotes the derivative with respect
to the variable $z$. These equations are easy to integrate, 
at least formally at 
the moment, by writing the conformal factor of the metric as
\be
e^{A(z)} = \left(-F^{\prime}(z)\right)^{N \over N + 4(D-2)} \ ,
\ee
in terms of some (yet unknown) function $F(z)$. Then, one easily finds that 
the scalar fields are given by
\be
e^{2\beta_i(z)} = {\left(-F^{\prime}(z)\right)^{\Delta /N} \over F(z) -b_i} ~, 
\ee
where $b_i$ are some integration constants (moduli) for all $i = 1, 2, \cdots , N$
and the exponent is $\Delta = 4N(D-2)/(N+4(D-2))$. 
Thus, according to this substitution, all it remains open 
is the differential equation for finding the suitable function $F(z)$.
This can be determined by appealing to the constraint \eqn{constr} among the 
fields $\beta_i$. Using their exponential form above, we arrive by multiplication
to the non-linear equation
\be
\left(-F^{\prime}(z)\right)^{\Delta} = \prod_{i=1}^{N} \left(F(z) - b_i \right)\equiv f
\label{maseq}\ ,
\ee
which has to be solved in all cases of interest.
  
The method we developed so far is quite general. For application to theories
of gauged supergravity we confine ourselves to the following cases for the
dimensionality of the spacetime $D$ and the number of scalar fields $N$: 
$(D, N) = (4, 8), ~ (5, 6)$ and $(7, 5)$. Note that in all three case the 
exponent of the corresponding non-linear differential equation is 
$\Delta = 4$. Of course, appropriate boundary conditions will be introduced
later in the integration of the master equation \eqn{maseq}. Also, the
range of $z$ has to be such that $F(z) \geq b_{\rm max}$ (the maximum value
of the moduli $b_i$) in order for the scalar fields $\beta_i$ to assume
real values.

Note at this point that had we chosen to work with the form of the
metric \eqn{mean1}, using the variable $r$ instead of $z$, a similar
ansatz would work for expressing the domain wall solutions in terms of an 
unknown function $F(r)$, namely
\be
e^{A(r)} = \left(\dot{F}(r)\right)^{N \over 4(D-2)} ~, ~~~~~ 
e^{2 \beta_i (r)} = {\dot{F}(r) \over F(r) - b_i} ~, 
\ee
where dot denotes the derivative with respect to $r$ and $b_i$ denote 
the appropriate moduli of integration. In this case we
easily find $(-F^{\prime}(z))^{\Delta} = (\dot{F}(r))^{N}$, and so the 
corresponding non-linear differential equation for $F(r)$ reads
\be
\left(\dot{F}(r)\right)^{N} = \prod_{i=1}^{N}\left(F(r)-b_i\right) . 
\label{nmaseq}
\ee
 
There are two advantages in using the variable $z$ instead of $r$.
First, the spectrum of the transverse traceless 
graviton fluctuations, obeying the massless scalar field equation, 
\be
\Phi(x, z) = {\rm exp}(i k \cdot x) {\rm exp}\left(-{D-2 \over 2} A(z)
\right) \Psi(z) ~, 
\ee
which represent plane waves propagating along the $(D-2)$-brane with a
$z$-dependent amplitude, can be cast directly into a time-independent
Schr\"odinger problem
\be
-\Psi^{\prime \prime}(z) + V(z) \Psi(z) = M^2 \Psi(z) ~, 
\label{susyqm1}
\ee
where $M^2 = -k \cdot k$. The corresponding Schr\"odinger potential
assumes the form 
\be
V(z) = W^2(z) + W^{\prime}(z)~; ~~~~~ W(z) \equiv {D-2 \over 2} 
A^{\prime} (z)
\label{susyqm2} 
\ee
and hence enjoys all properties of supersymmetric quantum mechanics.
Second, as we will see next, it is possible to treat the three cases
$(D, N)$ that are of interest in gauged supergravity all at once, 
because the value of the exponent $\Delta = 4$ is universal. Then, starting
from $(D,N)=(4, 8)$ we will obtain results for the other cases too by a simple
limiting procedure applied to the algebraic classification of the 
corresponding domain wall solutions.

\section{Algebraic classification}
\setcounter{equation}{0}
   
The non-linear differential equation \eqn{maseq} with $\Delta = 4$ 
can be viewed, when it is extended to the 
complex domain, as a Christoffel--Schwarz transformation from a 
closed polygon in the $z$-plane onto the upper-half $F$-plane. In this 
case the perimeter of the polygon is mapped to the real $F$-axis, 
whereas its vertices are mapped to points parametrized by the
moduli $b_i$. It is convenient to start from the case $(D, N) = (4, 8)$
by considering a closed octagon and the Christoffel--Schwarz 
transformation
\be
{dz \over dF} = (F-b_1)^{-\varphi_1 / \pi} (F-b_2)^{-\varphi_2 / \pi}
\cdots (F-b_8)^{-\varphi_8 / \pi} \ ,
\ee
making the canonical choice
\be
\varphi_1 = \varphi_2 = \cdots = \varphi_8 = {\pi \over 4} \ ,
\ee
for generic values of the moduli $b_i$. Letting 
\be
x =F(z) ~, ~~~~~ y =F^{\prime}(z) ~,
\ee
we arrive at the algebraic curve
\be
y^4 = (x-b_1) (x-b_2) \cdots (x-b_8) ~, 
\ee
which has to be {\em uniformized} in terms of a complex variable, 
say $u$, as $x = x(u)$ and $y = y(u)$. This is a difficult problem 
in practice 
for generic values of the moduli $b_i$, but as soon as this is done
we may obtain $u(z)$ by inverting the solution of the equation
\be
{dz \over du} = {1 \over y(u)}{dx(u) \over du} 
\label{bouco}
\ee
and hence construct the desired solution of the master equation
that determines the domain walls of maximal gauged supergravity
in four spacetime dimensions according to the ansatz (2.9) and (2.10).   

It is clear that in the corner of the moduli space, where
all $b_i$ become equal to each other, the scalar fields vanish and
the geometry of the metric coincides with that of $AdS_4$, as it should. 
In fact, this solution provides the boundary condition for integrating
the equation \eqn{bouco}, and hence determine the domain wall 
solutions for generic points of the moduli space, namely, we demand
that the domain wall approaches the trivial $AdS_4$ solution as
$F \rightarrow \infty$ (or equivalently letting $z \rightarrow 0^+$,  
which fixes the origin). Appropriate restrictions also have to be 
introduced so that the resulting solutions for the scalar fields
and the conformal factor of the metric turn out to be real, despite
the original formulation of the Christoffel--Schwarz transformation
in the complex domain.  

Standard techniques from algebraic geometry yield the
classification of the algebraic curves in table 1 below
(written in irreducible form),
which describe the domain walls in $D=4$, $N=8$ according to genus.
We also note that the case where the unbroken symmetry group equals
the Cartan subgroup of $SO(8)$, i.e. $SO(2)^4$, it corresponds to the 
extremal supersymmetric limit of the most general rotating M2-brane 
solution which depends on 
four rotating parameters. In general, the same remark
holds true for the tables corresponding to all $p$-branes 
that will be listed later 
in the paper. Namely, when the 
original symmetry group $SO(N)$ is broken down to its Cartan subgroup, the 
corresponding solution is the extremal supersymmetric limit of the most 
general rotating $p$-brane solution, with the number of its
rotating parameters equal 
to the dimension of the Cartan subgroup of $SO(N)$.

\begin{center}
\begin{tabular}{c|l|l}
{\hskip-3pt \em Genus} & {\em Irreducible Curve} & {\em Isometry Group}\\
\hline
 9 & $y^4=(x-b_1)(x-b_2) \cdots (x-b_7)(x-b_8)$ & None \\
\hline
 7 & $y^4=(x-b_1)(x-b_2) \cdots (x-b_6)(x-b_7)^2$ & $SO(2)$ \\
\hline
 6 & $y^4=(x-b_1)(x-b_2) \cdots (x-b_5)(x-b_6)^3$ & $SO(3)$ \\
\hline
 5 & $y^4=(x-b_1) \cdots (x-b_4)(x-b_5)^2(x-b_6)^2$ 
& $SO(2) \times SO(2)$ \\
\hline
 4 & $y^4=(x-b_1)(x-b_2)(x-b_3)(x-b_4)^2(x-b_5)^3$ 
& $SO(2) \times SO(3)$ \\
\hline
 3 & $y^4=(x-b_1) \cdots (x-b_4)(x-b_5)^4$ & $SO(4)$ \\
    & $y^4=(x-b_1)(x-b_2)(x-b_3)(x-b_4)^5$ & $SO(5)$ \\ 
    & $y^4=(x-b_1)(x-b_2)(x-b_3)^3(x-b_4)^3$ 
& $SO(3) \times SO(3)$ \\
    & $y^4=(x-b_1)(x-b_2)(x-b_3)^2(x-b_4)^2(x-b_5)^2$ 
& $SO(2) \times SO(2) \times SO(2)$ \\
\hline
 2 & $y^4 = (x-b_1)(x-b_2)^2(x-b_3)^2(x-b_4)^3$ 
& $SO(2) \times SO(2) \times SO(3)$ \\
\hline
 1 & $y^4=(x-b_1)(x-b_2)(x-b_3)^6$ & $SO(6)$ \\
    & $y^4= (x-b_1)(x-b_2)(x-b_3)^2(x-b_4)^4$ 
& $SO(2) \times SO(4)$ \\
    & $y^4=(x-b_1)(x-b_2)^2(x-b_3)^5$ & $SO(2) \times SO(5)$ \\
    & $y^4=(x-b_1)^2(x-b_2)^3(x-b_3)^3$ 
& $SO(2) \times SO(3) \times SO(3)$ \\
    & $y^2=(x-b_1)(x-b_2)(x-b_3)(x-b_4)$ 
& $SO(2)^4$ \\
\hline
 0 & $y^4=(x-b_1)(x-b_2)^7$ & $SO(7)$ \\
    & $y = (x-b)^2$ & $SO(8)$ \hskip .5cm (Maximal) \\
    & $y^2=(x-b_1)(x-b_2)^3$ & $SO(2) \times SO(6)$\\
    & $y^4=(x-b_1)(x-b_2)^3(x-b_3)^4$ & $SO(3) \times SO(4)$ \\
    & $y^4=(x-b_1)^3(x-b_2)^5$ & $SO(3) \times SO(5)$ \\
    & $y = (x-b_1)(x-b_2)$ & $SO(4) \times SO(4)$\\
    & $y^2=(x-b_1)(x-b_2)(x-b_3)^2$ 
& $SO(2)^2 \times SO(4)$\\
\hline 
\end{tabular}
\end{center}
\begin{center} 
Table 1: Curves and symmetry groups of domain walls in $D=4$, $N=8$ 
supergravity.
\end{center}

The symmetry groups refer to the special regions of the moduli space
where some of the parameters $b_i$ are allowed to coincide, thus 
lowering the genus of the corresponding Riemann surfaces, which in
turn lead to simplifications in the domain wall solutions as some
of the scalar fields are linearly related to others.

The algebraic classification of the domain walls in $D=5$ supergravity
with $N=6$ follows immediately from the list above by considering only
those solutions with $SO(2)$ isometry and letting the two 
coalescing moduli tend to infinity. Then, the non-linear differential
equation \eqn{maseq} with $N=8$ becomes (after appropriate rescaling) 
the corresponding 
equation with $N=6$, which is appropriate for $D=5$ supergravity.
In the Christoffel--Schwarz transformation this amounts to 
degenerating the closed octagon by letting two of its 
vertices coincide and mapping the resulting double vertex to 
infinity. Thus,
the classification presented in table 2 follows immediately:
 
\begin{center}
\begin{tabular}{c|l|l}
{\em Genus} & {\em Irreducible Curve} & {\em Isometry Group}\\
\hline
 7 & $y^4=(x-b_1)(x-b_2) \cdots (x-b_5)(x-b_6)$ & None \\
\hline
 5 & $y^4=(x-b_1)(x-b_2)(x-b_3)(x-b_4)(x-b_5)^2$ & $SO(2)$ \\
\hline
 4 & $y^4=(x-b_1)(x-b_2)(x-b_3)(x-b_4)^3$ & $SO(3)$ \\
\hline
 3 & $y^4=(x-b_1)(x-b_2)(x-b_3)^2(x-b_4)^2$ 
& $SO(2) \times SO(2)$ \\
\hline
 2 & $y^4=(x-b_1)(x-b_2)^2(x-b_3)^3$ 
& $SO(2) \times SO(3)$ \\
\hline
 1 & $y^4=(x-b_1)(x-b_2)(x-b_3)^4$ & $SO(4)$ \\
    & $y^4=(x-b_1)(x-b_2)^5$ & $SO(5)$ \\ 
    & $y^4=(x-b_1)^3(x-b_2)^3$ 
& $SO(3) \times SO(3)$ \\
    & $y^2=(x-b_1)(x-b_2)(x-b_3)$ 
& $SO(2)^3$ \\
\hline
 0 & $y^2 = (x-b_1)(x-b_2)^2$ 
& $SO(2) \times SO(4)$ \\
    & $y^2=(x-b)^3$ & $SO(6)$ \hskip .5cm (Maximal) \\
\hline 
\end{tabular}
\end{center}
\begin{center}
Table 2: Curves and symmetry groups of domain walls in 
$D=5$, $N=6$ supergravity.
\end{center}

Likewise, letting three vertices first coincide and then mapping
them to infinity, which amounts to factoring out an $SO(3)$ 
isometry, we obtain table 3 for the algebraic classification of
the domain walls in $D=7$ supergravity with $N=5$: 

\begin{center}
\begin{tabular}{c|l|l}
{\em Genus} & {\em Irreducible Curve} & {\em Isometry Group}\\
\hline
 6 & $y^4=(x-b_1)(x-b_2)(x-b_3)(x-b_4)(x-b_5)$ & None \\
\hline
 4 & $y^4=(x-b_1)(x-b_2)(x-b_3)(x-b_4)^2$ & $SO(2)$ \\
\hline
 3 & $y^4=(x-b_1)(x-b_2)(x-b_3)^3$ & $SO(3)$ \\
\hline
 2 & $y^4=(x-b_1)(x-b_2)^2(x-b_3)^2$ & $SO(2)^2$ \\
\hline
 1 & $y^4=(x-b_1)^2(x-b_2)^3$ & $SO(2) \times SO(3)$  \\
\hline
 0 & $y^4=(x-b_1)(x-b_2)^4$ & $SO(4)$ \\
    & $y^4=(x-b)^5$ & $SO(5)$ \hskip .5cm (Maximal) \\ \hline
\end{tabular}
\end{center}
\begin{center}
Table 3: Curves and symmetry groups of domain walls in $D=7$, $N=5$ 
supergravity.
\end{center}

Two technical remarks are in order before proceeding further.
First, suppose that a moduli $b$ of the Christoffel--Schwarz 
transformation
underlying \eqn{maseq} is taken to infinity. 
This will lead to elimination of the corresponding factor
$(F-b)^{\gamma}$ from the algebraic curve, where $\gamma$ denotes the 
associated degree of degeneracy of the vertex. 
It is practically achieved by first
rescaling $z$ to a new variable 
$z^{\prime} = z(-b)^{\gamma / \Delta}$ and then letting  
$-b \rightarrow \infty$. This actually amounts to rescaling the 
coupling constant $g^2$ by a factor $(-b)^{\gamma / \Delta}$,
which was previously set equal to 1 for
simplicity, but it can be reinstated any moment in the various
equations. We will see later that this limit provides a smearing
for the various 
brane distributions which allows to construct, 
using also various $U$-duality transformations,
solutions representing $p$-brane distributions for all values 
of $p$ in string theory.
For the moment,
it is sufficient to justify the classification presented in
tables 2 and 3 following the complete list of table 1. We note,      
however, that in all three cases of interest in gauged supergravity 
the associated polygon in the $z$ plane is closed for generic
values of the moduli $b_i$; it has 8, 7 and 6 verices for 
$(D, N) = (4, 8)$, $(5, 6)$ and $(7, 5)$, respectively. Of course,
at certain corners of the moduli space, where it degenerates further, 
the polygon may turn open for appropriate large isometry groups 
(see, for instance, table 1 for
solutions with $SO(n\geq4)$ factors). 

Second, the interpretation of the differential equation
\eqn{nmaseq} as a Christoffel--Schwarz transformation in terms of
the variable $r$ differs from the interpretation given 
to equation \eqn{maseq}
in terms of the variable $z$ in that we have to consider the mapping
of an open polygon in the $r$ plane onto the upper-half $F$-plane.
Namely, instead of an octagon at generic points, 
what we are considering now is an 
open polygon with $N$ vertices each one having $\pi / N$ as defect
angle, and another vertex pulled at infinity in the $r$-plane, which
necessarily has defect angle $\pi$; furthermore, 
the latter is mapped to infinity
in the $F$-plane. Therefore, changing variables
from $r$ to $z$ is expected to be transcendental at generic points
of the moduli space; put differently, the genus of the algebraic
curves based on \eqn{maseq} or \eqn{nmaseq} will not be the same
at generic points of the moduli space. It should be realized
that this is not a problem as in the corresponding uniformization 
of the surfaces there are different multiple coverings along the branch 
cuts. At certain degenarate limits, however, where
the isometry group of the solutions is appropriately chosen, the
bounded regions in the $z$- or $r$-plane may have the same shape 
and hence the genus of the corresponding curves will be equal. 
Presently, we have chosen to work with the
$z$-parametrization instead of $r$ 
for the two main reasons that were explained in section 2.

\section{Distributions of M2-, M5- and D3-branes}
\setcounter{equation}{0}

It was shown in the case of maximally gauged supergravity 
in $D=7$, 4 and 5 dimensions that 
the construction of domain walls gives 
rise to particular solutions of supergravity in higher dimensions, which
describe the field theory limit of 
a large number of M5-, M2- and D3-branes 
distributed in various hypersurfaces embedded in the $N$-dimensional 
space transverse to the branes.
In particular, the higher dimensional 
metrics for the various distributions of branes have the form
\be
{\rm M5\!-\! brane}:\qq 
ds^2 = H_0^{-1/3} \eta_{\m\n} dx^\m dx^\n + H_0^{2/3} (dy_1^2+dy_2^2+\dots +
dy_5^2)\ ,
\label{M55}
\ee
\be
{\rm M2\!-\! brane}:\qq 
ds^2 = H_0^{-2/3} \eta_{\m\n} dx^\m dx^\n + H_0^{1/3} (dy_1^2+dy_2^2+\dots +
dy_8^2)\ ,
\label{M22}
\ee
and 
\be
{\rm D3\!-\! brane}:\qq 
ds^2 = H_0^{-1/2} \eta_{\m\n} dx^\m dx^\n + H_0^{1/2} (dy_1^2+dy_2^2+\dots +
dy_6^2)\ .
\label{D33}
\ee
In all cases $H_0$ is a harmonic function 
in the $N$-dimensional space $\IR^N$
transverse to the brane parametrized by the coordinates $y_i$ and 
it is given by
\be
H_0^{-1}  = {4 \ov R^4}  f^{1/2} \sum_{i=1}^N {y_i^2\ov (F-b_i)^2}\ ,
\label{dhj1}
\ee
where $f$ is defined in \eqn{maseq}.
The coordinate $F$ is determined in terms of the transverse coordinates
$y_i$ as a solution of the algebraic equation
\be
\sum_{i=1}^N  {y_i^2\ov F-b_i} =4 g^{D-5}\ .
\label{jk4}
\ee

The algebraic equation \eqn{jk4} for $F$ cannot be solved 
analytically for general choices of the constants $b_i$.
However, this becomes practically possible when some of the $b_i$'s coincide
in such a way that the
degree of the corresponding algebraic equations with respect 
to $F$ is reduced to 4 or less. 
It can also be shown that $H_0$, as defined in \eqn{dhj1} and \eqn{jk4}, 
is a harmonic function in $\IR^N$, as it should. 
We may solve this constraint by introducing the change of coordinates
\be
y_i = 2 g^{(D-5)/2} (F-b_i)^{1/2} \hat x_i\ ,\qq i=1,2,\dots , N\ ,
\label{ejh1}
\ee
where the $\hat x_i$'s define a unit $N$-sphere.
Then, the $N$-dimensional flat metric
in the transverse part of the brane metric \eqn{M55}--\eqn{D33} 
can be written as
\be
 \sum_{i=1}^N dy_i^2 = g^{D-5} \sum_{i=1}^N {\hat x_i^2\ov F-b_i}\ dF^2
+4 g^{D-5} \sum_{i=1}^N (F-b_i) d\hat x_i^2\ .
\label{dm22}
\ee
The metrics \eqn{M55}--\eqn{D33}
become asymptotically $AdS_D\times S^{N-1}$ for large radial distances,
with $D$ and $N$ taking their appropriate values. The radius of the sphere  
is always $R$, whereas for $AdS_D$ it is  
$(D-3) R/2$.

We note for completeness that brane solutions which are asymptotically flat
can be obtained by replacing $H_0$ in \eqn{M55}--\eqn{D33} by $H=1+H_0$. 
Then, in this context, the length parameter 
$R$ has a microscopic interpretation using 
the eleven-dimensional Planck scale $l_{\rm P}$ or the string scale 
$\sqrt{\a'}$ and the string coupling $g_s$, 
and the number of branes $N_{\rm b}$, which is assumed large.
For M5-branes we have
$R^3 =  \pi N_{\rm b} l_{\rm P}^3$, for M2-branes 
$R^6=32 \pi N_{\rm b} l_{\rm P}^6$, whereas for D3-branes 
we have $R^4= 4\pi N_{\rm b} g_s \a'^2$.

\section{Dualities, smearing and $p$-brane distributions }
\setcounter{equation}{0}

In this section we develop a 
smearing procedure which allows to construct brane 
distributions for all $p$-branes of the type-II string theory. 
Starting 
with the M-theory branes we immediately obtain solutions 
corresponding to distributions of fundamental strings NS1 and D4-branes of
type-IIA by simply dimensionally 
reducing the M2- and M5-brane solutions, respectively,
along one of the $x^\m$-coordinates.
An S-duality transformation gives from the NS1 configurations
(within its type-IIB interpretation) a solution
representing a distribution of D1-branes.
The solution representing a distribution of fundamental NS1 strings is given, in
the string frame, by
\ba
ds^2 &=& H^{-1}(-dt^2 + dx_1^2) + dy_1^2+\dots dy_8^2\ ,
\nonumber\\
B_{01} & =&  H^{-1}\ ,\qq e^{-2 \Phi} \ =\  H\ .
\label{ns11}
\ea
The harmonic function $H$ is exactly the same as for the M2-branes 
and the corresponding curves and symmetry groups are 
given as in table 1.
For D1- and D4-branes the corresponding metrics (in the string frame) 
and dilaton fields, omitting the 
associated antisymmetric tensor which is also turned on, are given by
\ba
ds^2 & = & 
H^{-1/2} \eta_{\m\n} dx^\m dx^\n + H^{1/2} (dy_1^2+dy_2^2+\dots +
dy_{9-p}^2)\ ,
\nonumber\\
e^{4\Phi} &= & H^{3-p} \ ,
\label{ddbr}
\ea
with $p=1$ and $p=4$, respectively. 
Again, the harmonic function $H$ is exactly the same as for the M2-
and M5-branes, respectively, and the corresponding curves
and symmetry groups are given as in tables 1 and 3. 

However, in order to dimensionally reduce along a direction which is
transverse to the M-branes, we have to employ a smearing procedure. 
Recall first that for 
single-centered solutions, the smearing procedure amounts to simply 
(re)distributing the branes along an infinite line identified with one of the 
transverse directions. In this way, the relevant harmonic function becomes 
independent of the corresponding coordinate, which in turn 
allows to perform the 
dimensional reduction. 
However, in our case this procedure is not applicable as it stands.
Instead, we consider the following limit,
\be
b_N= -\e^{-2N} g^{3-D}\ ,
\qq x^\m\to x^\m \e^{N\ov 2(D-2)}\ ,\qq g\to g\e\ ,\qq
y_i\to \ y_i \e^{D-5\ov 2}\ ,
\label{lii}
\ee
where $\e\to 0$ is an auxiliary dimensionless infinitesimal quantity. 
We note the explicit appearance of the factor $g^{3-D}$ in the expression 
for $b_N$ above inserted on purely dimensional grounds,
since the constants $b_i$
have dimension of $({\rm length})^{D-3}$.
This limit is well defined; one easily sees from \eqn{ejh1}
that $\hat x_N = {\cal O}(\e^N)$,
whereas the rest of the unit vectors stay finite and define an $(N-1)$-sphere.
It follows that, in this limit, the dependence on the coordinate
$y_N$ disappears and hence we may consider the 
dimensional reduction of the 
M-theory brane solutions \eqn{M55} and \eqn{M22} as before. 
It is worth emphasizing that the smearing based on \eqn{lii}
holds true for all three cases of section 4 (and only for these).

Starting first from the M2-brane solution,
we obtain after dimensional reduction along the
direction  transverse to the brane a solution representing a 
distribution of D2-branes.
The metric and dilaton are given by \eqn{ddbr} with $p=2$, where the 
harmonic function $H$ is given by \eqn{dhj1} and \eqn{jk4} 
with $N=7$ and $D=4$,
whereas the corresponding curves and 
symmetry groups are given in table 4 below.

%\newpage

%%%%%%%%%%%%%%%%%%%%%%%%%%%%%%%%%%%%%%%%%%%%%%%%%%%%%%%%%%%%%%

\begin{center}
\begin{tabular}{c|l|l}
{\hskip-3pt \em Genus} & {\em Irreducible Curve} & {\em Isometry Group}\\
\hline
 9 & $y^4=(x-b_1)(x-b_2) \cdots (x-b_6)(x-b_7)$ & None \\
\hline
 7 & $y^4=(x-b_1)(x-b_2) \cdots (x-b_5)(x-b_6)^2$ & $SO(2)$ \\
\hline
 6 & $y^4=(x-b_1)(x-b_2) \cdots (x-b_4)(x-b_5)^3$ & $SO(3)$ \\
\hline
 5 & $y^4=(x-b_1) \cdots (x-b_3)(x-b_4)^2(x-b_5)^2$ 
& $SO(2) \times SO(2)$ \\
\hline
 4 & $y^4=(x-b_1)(x-b_2)(x-b_3)^2(x-b_4)^3$ 
& $SO(2) \times SO(3)$ \\
\hline
 3 & $y^4=(x-b_1) \cdots (x-b_3)(x-b_4)^4$ & $SO(4)$ \\
    & $y^4=(x-b_1)(x-b_2)(x-b_3)^5$ & $SO(5)$ \\ 
    & $y^4=(x-b_1)(x-b_2)^3(x-b_3)^3$ 
& $SO(3) \times SO(3)$ \\
    & $y^4=(x-b_1)(x-b_2)^2(x-b_3)^2(x-b_4)^2$ 
& $SO(2)^3$ \hskip .5cm \\
\hline
 2 & $y^4 = (x-b_1)^2(x-b_2)^2(x-b_3)^3$ 
& $SO(2) \times SO(2) \times SO(3)$ \\
\hline
 1 & $y^4=(x-b_1)(x-b_2)^6$ & $SO(6)$ \hskip .5cm \\
    & $y^4= (x-b_1)(x-b_2)^2(x-b_3)^4 $ 
& $SO(2) \times SO(4)$ \hskip .5cm \\
& $y^4=(x-b_1)^2 (x-b_2)^5$ &$SO(2) \times SO(5)$  \hskip .5cm \\
\hline
 0     & $y^4=(x-b_1)^3 (x-b_2)^4$ &$SO(3) \times SO(4)$  \hskip .5cm \\
       &  $y^4 = (x-b)^7$ & $SO(7)$  (Maximal) \\
\hline 
\end{tabular}
\end{center}
\begin{center} 
Table 4: Curves and symmetry groups of D2-brane distributions.
\end{center}

%\newpage

After the limiting procedure \eqn{lii}  
the dimensional reduction of the M5-brane solution 
along a transverse direction gives a solution representing a 
distribution of NS5-branes in type-IIA string theory.
The string frame metric, the antisymmetric tensor field strength 
and dilaton field are
\ba
ds^2 &=& -dt^2 + dx_1^2 +\dots + dx_5^2 + H (dy_1^2+\dots dy_4^2)\ ,
\nonumber\\
e^{2 \Phi} & = & H\ ,\qq H_{ijk} \ =\ \e_{ijkl} \del_l H\ ,\quad i=1,2,3,4\ ,
\label{ns55}
\ea
where the harmonic function is given by \eqn{dhj1} and \eqn{jk4}
with $N=4$ and 
$D=7$. After a T-duality along a direction parallel to the NS5-branes,
so that a solution of type-IIB emerges,
we can apply an S-duality transformation in order to obtain 
a solution representing 
a distribution of D5-branes 
with metric and dilaton given by \eqn{ddbr} with $p=5$. 
For both NS5 and D5 cases, the curves and 
symmetry groups are given in table 5 below.

%%%%%%%%%%%%%%%%%%%%%%%%%%%%%%%%%%%%%%%%%%%%%%%%

\begin{center}
\begin{tabular}{c|l|l}
{\em Genus} & {\em Irreducible Curve} & {\em Isometry Group}\\
\hline
 3 & $y^4=(x-b_1)(x-b_2)(x-b_3)(x-b_4)$ & None \\
\hline
 1 & $y^4=(x-b_1)(x-b_2)(x-b_3)^2$ & $SO(2)$ \\
\hline
 0 & $y^4=(x-b_1)(x-b_2)^3$ & $SO(3)$ \\
    & $y^2=(x-b_1)(x-b_2)$ & $SO(2)\times SO(2)$   \\ 
    & $y=(x-b)$ & $SO(4)$ (maximal) \\
\hline 
\end{tabular}
\end{center}
\begin{center} 
Table 5: Curves and symmetry groups of NS5- and D5-brane 
distributions.
\end{center}

%\newpage

%%%%%%%%%%%%%%%%%%%%%%%%%%%%%%%%%%%%%%%%%%%%%%%%%%%%%%%

Finally, we may apply a similar smearing procedure to the D5-brane
distributions of table 5. After a T-duality transformation,
we obtain solutions representing distributions of D6-branes. The 
harmonic function is given again by \eqn{dhj1} and \eqn{jk4}
with $N=3$ and $D=7$. The corresponding curves and symmetry groups are 
given in table 6.

\begin{center}
\begin{tabular}{c|l|l}
{\em Genus} & {\em Irreducible Curve} & {\em Isometry Group}\\
\hline
 3 & $y^4=(x-b_1)(x-b_2)(x-b_3)$ & None \\
\hline
 1 & $y^4=(x-b_1)(x-b_2)^2$ & $SO(2)$   \\
\hline
 0 & $y^4=(x-b_1)^3$ & $SO(3)$ (maximal) \\
\hline 
\end{tabular}
\end{center}
\begin{center} 
Table 6: Curves and symmetry groups of D6-brane 
distributions.
\end{center}

%\bigskip 
%\begin{center}
%\begin{tabular}{c|l|l}
%{\em Genus} & {\em Irreducible Curve} & {\em Isometry Group}\\
%\hline
% 1 & $y^4=(x-b_1)(x-b_2)$ & None \\
%\hline
%  0 & $y^2=(x-b_1)$ & $SO(2)$ (maximal) \\
%\hline 
%\end{tabular}
%\end{center}
%\begin{center} 
%Table 7: Curves and symmetry groups of D7-brane 
%distributions
%\end{center}

%\newpage 

In this section we considered so far the limiting procedure in terms of
M or string theory solutions. Similar considerations can be made in 
terms of lower dimensional theories of gauged supergravity, as in 
\cite{cvepope} for reductions on certain singular limits of $S^4$.

\section{An example of a distribution of NS5-branes}
\setcounter{equation}{0}

In the case of NS5-branes (or D5-branes) it is possible to explicitly solve the
quartic equation \eqn{jk4} (with $N=4$ and $D=7$)
for $F$ and substitute the result back into \eqn{dhj1} in order to obtain
an explicit expression for the corresponding 
harmonic function. However, the resulting 
expression is not very illuminating to present in detail for general 
values of the moduli.

We focus attention to distributions of NS5-branes, where
the constant $R$ (in analogy to previous cases) has a microscopic 
interpretation in terms of the number $N$ of NS5-branes and the string scale 
$\a'$ as $R^2= N \a'$. 
For genus zero, besides the solution with isometry $SO(4)$, we may explicitly 
present the solution \eqn{ns55}. Recall that 
the solution with symmetry $SO(2)\times SO(2)$
has already been given in \cite{sfe1} and it represents the field of 
a large number of NS5-branes 
uniformly distributed on a circle. In that case it was shown that there is an
{\it exact} conformal field theory corresponding to the coset model 
$SL(2,\IR)/\IR \times SU(2)/U(1)$ with level equal to the number 
$N$ of NS5-branes. 

Here, we present for completeness the other case corresponding to 
the genus zero surface that preserves an $SO(3)$ symmetry, 
according to table 5. 
For this, it is convenient to use a basis for the unit vectors 
$\hat x_i$ that define the three-sphere in such a way that it is in one to one
correspondence with the decomposition of 
the vector representation $\bf 4$ of $SO(4)$ with respect to the subgroup
$SO(3)$, as ${\bf 4}\to {\bf 3} \oplus {\bf 1}$. 
Hence, we choose 
\be
\pmatrix{\hat x_1\cr \hat x_2}\ = \  
\cos\th \sin\om \pmatrix{\cos\varphi\cr\sin\varphi} \ ,
\quad 
\hat x_3 \ =\  \cos\th \cos\om\ , \quad \hat x_4 \ =\  \sin\th\ .
\label{jwoi1}
\ee
It is also convenient to choose the constants $b_i$ as follows 
\be
b_1=b_2=b_3=0\ ,\qq b_4=-l^2\ ,
\label{fdj1}
\ee   
where $l$ is a real constant. 
The expressions following \eqn{ns55} 
for the four-dimensional transverse part of the metric, 
the antisymmetric tensor and the dilaton fields are given explicitly by 
\ba
ds^2 & = & {1\ov 4} \left(1+{l^2\ov r^2}\right)^{1/2}
 \left({dr^2\ov r^2 + l^2}
+d\th^2 +{r^2 \cos^2\th\ov r^2 +l^2 \cos^2\th}(d\om^2 +\sin^2\om d\varphi^2)
\right)\ ,
\nonumber\\
B_{\om\varphi} & = & {1\ov 4} \sin\om \left(\th + {r^2 \cos\th \sin\th
\ov r^2 +l^2 \cos^2\th}\right)\ ,\qq
e^{2\Phi} \ = \ {\left(1+{l^2\ov r^2}\right)^{1/2}\ov r^2+l^2\cos^2\th}\ .
\label{jds11}
\ea
In this case, the distribution of NS5-branes is taken 
over a segment of length $2l$ 
along the $x_4$-axis with center at the origin. The location of the 
brane distribution manifests as a naked curvature singularity at
$r=0$ of the metric in \eqn{jds11}. Note also that the 
analytic continuation $l^2\to -l^2$ yields a naked singularity
at $r=l$ corresponding to a distribution of NS5-branes on a three-sphere.

It will be interesting to know whether there is an exact conformal
field theory corresponding to the background \eqn{jds11}.

\bigskip

\centerline{\bf Acknowledgments}
\no
One of us (I.B.) wishes to thank the organizers of the conference
for their kind invitation to present an account of the results. 
This research was supported by the European Union under contracts 
HPRN-CT-2000-00122 and -00131, as well as by the Swiss National Science
Foundation and the Swiss Office for Education and Science.

\end{document}